\documentclass[aps,prd,superscriptaddress,amsfonts,amssymb,amsmath,eqsecnum,nofootinbib,twocolumn,floatfix]{ revtex4} 

 \usepackage{color,graphicx}
 \usepackage[utf8]{inputenc}
 \usepackage[T2A]{fontenc}
 \usepackage[russian,english]{babel}
 \definecolor{darkblue}{rgb}{0,0,0.7}
\definecolor{darkred}{rgb}{0.7,0,0}
\definecolor{darkgreen}{rgb}{0,0.4,0}
\usepackage[unicode, colorlinks, citecolor=darkblue, linkcolor=darkred, urlcolor=blue]{hyperref}
\allowdisplaybreaks
\begin{document}

\author{Albert Nazmiev} 
\email{nazmiev.ai15@physics.msu.ru}
\affiliation{Faculty of Physics, M.V. Lomonosov Moscow State University, Leninskie Gory, Moscow 119991,  Russia}
\author{Sergey P. Vyatchanin}
\affiliation{Faculty of Physics, M.V. Lomonosov Moscow State University, Leninskie Gory, Moscow 119991,  Russia}
\affiliation{Quantum Technology Centre, Moscow State University, Moscow 119991 Russia}
\date{\today}
	
\title{Stable Optical Rigidity Based on Dissipative Coupling}

\begin{abstract}
We show that the stable optical rigidity can be obtained in a Fabry-Perot cavity with dissipative optomechanical coupling and with detuned pump, corresponding conditions are formulated. An optical detection of a weak classical mechanical force with usage of this rigidity is analyzed. The sensitivity of small force measurement can be better than the standard quantum limit (SQL).
\end{abstract}

\maketitle

\section{Introduction}

Resonant optomechanics \cite{AspelmeyerRMP2014} investigates interaction between an optical cavity and a free mass or a
mechanical oscillator. The simplest optomechanical interaction is based on the radiation pressure effect in
which a force proportional to optical power or number of the optical quanta, circulating in a 1D optical cavity, acts on a
test mass so that the size of the optical cavity increases with increase of number of the optical quanta localized in there. 
Such interaction is usually called as dispersive coupling. Systems having several 
degrees of freedom allow more complex optomechanical interactions, including radiation pulling (negative radiation 
pressure) \cite{Povinelli05ol,maslov13pra}, optomechanical interaction proportional to the quadrature of 
electromagnetic field \cite{93a1VyMaJETP,96a1VyMaJETP,matsko97apb,02a1KiLeMaThVyPRD} and the interaction depending on 
the speed and not the coordinate of the mechanical system \cite{90BrKhPLA,00a1BrGoKhThPRD}.

Optomechanical interaction is important in precise measurements which use an efficient quantum transduction 
mechanism between the mechanical and optical degrees of freedom allowing various sensors, like gravitational wave 
detectors \cite{aLIGO2013, KAGRA2013, aLIGO2014, GEO2014, aLIGO2015,aVirgo2015, GWpro_2018, GW_2016b, 
GW_2017, GW_2017b, aVirgo2018}, torque sensors \cite{WuPRX2014}, and magnetometers \cite{ForstnerPRL2012}.

Sensitivity of the mechanical coordinate measurement in an optomechanical system usually is restricted by the so-called 
standard quantum limit (SQL) \cite{Braginsky68,BrKh92} due to the quantum backaction. The SQL was investigated in various 
configurations ranging from the macroscopic gravitational wave detectors \cite{02a1KiLeMaThVyPRD} to the microcavities 
\cite{Kippenberg08,DobrindtPRL2010}. Sensitivity of the other types of measurements being derivatives of the coordinate 
detection is also limited by the SQL.  An example of such measurement is detection of a classical force acting on a 
mechanical degree of freedom of an optomechanical system. However, the  SQL of the force measurement is not an  
unavoidable limit. Several approaches can surpass the SQL, for example, variational measurement \cite{93a1VyMaJETP, 
95a1VyZuPLA, 02a1KiLeMaThVyPRD}, optomechanical velocity measurement using dispersive coupling \cite{90BrKhPLA,00a1BrGoKhThPRD}, 
measurements in optomechanical systems with optical rigidity \cite{99a1BrKhPLA, 01a1KhPLA}.  Quantum speed meter based on dissipative
 coupling was proposed recently \cite{16a1PRAVyMa}.

Among variety of the optomechanical interactions the dissipative coupling takes a special place. The dissipative 
coupling is characterized by dependence of an optic cavity relaxation on a mechanical coordinate (in case of a Fabry-Perot 
cavity the mechanical coordinate changes transparency of the input mirror), whereas dispersive coupling is characterized by the 
dependence of a cavity frequency  on the coordinate. The system with dissipative coupling cannot be considered lossless 
anymore. But the dissipation here does not lead to decoherence or absorption of light, instead, it results 
in lossless coupling between a continuous optical wave and a mode of an optical cavity. The  cavity with 
dissipative coupling can be used as a perfect transducer between the continuous optical wave and the 
mechanical degree of freedom, allowing efficient cooling of the mechanical oscillator 
\cite{ElstePRL2009,LiPRL2009,WuPRX2014, HryciwOpt2015,SawadskyPRL2015},
exchange of the quantum states between the optical and mechanical degrees of freedom, mechanical squeezing 
\cite{KronwaldPRA2013, TanPRA2013,ZhuJAP2014,QuPRA2015}, and a combination of cooling and squeezing 
\cite{GuPRA2013,GuOE2013}. A combination of conventional, dispersive, and dissipative coupling adds more complexity to 
the interaction and leads to the new effects \cite{HuangPRA2010,HuangPRA2010b}.

Dissipative coupling was proposed theoretically \cite{ElstePRL2009} and implemented experimentally about ten years ago
\cite{LiPRL2009,WeissNJP2013,WuPRX2014, HryciwOpt2015}. It was investigated in different
optomechanical systems, including a Fabry-Perot interferometer \cite{LiPRL2009,WeissNJP2013,WuPRX2014, HryciwOpt2015}, 
a Michelson-Sagnac interferometer (MSI) \cite{XuerebPRL2011, TarabrinPRA2013, SawadskyPRL2015}, and the ring resonators 
\cite{HuangPRA2010,HuangPRA2010b}.

In this paper we report on one more important feature of a cavity with dissipative coupling. Such cavity, non-resonantly 
pumped, introduces a {\em stable} optical rigidity into the mechanical degree of freedom. Recall that the optical rigidity based 
on conventional dispersive coupling (in non-resolved side band case) is unstable and can be used only with feedback. We 
formulate the conditions of stability 
and show that the stable optical rigidity based on dissipative coupling allows to surpass the SQL.

\section{Hamiltonian approach}\label{model}

\begin{figure}
 \includegraphics[width=0.45\textwidth]{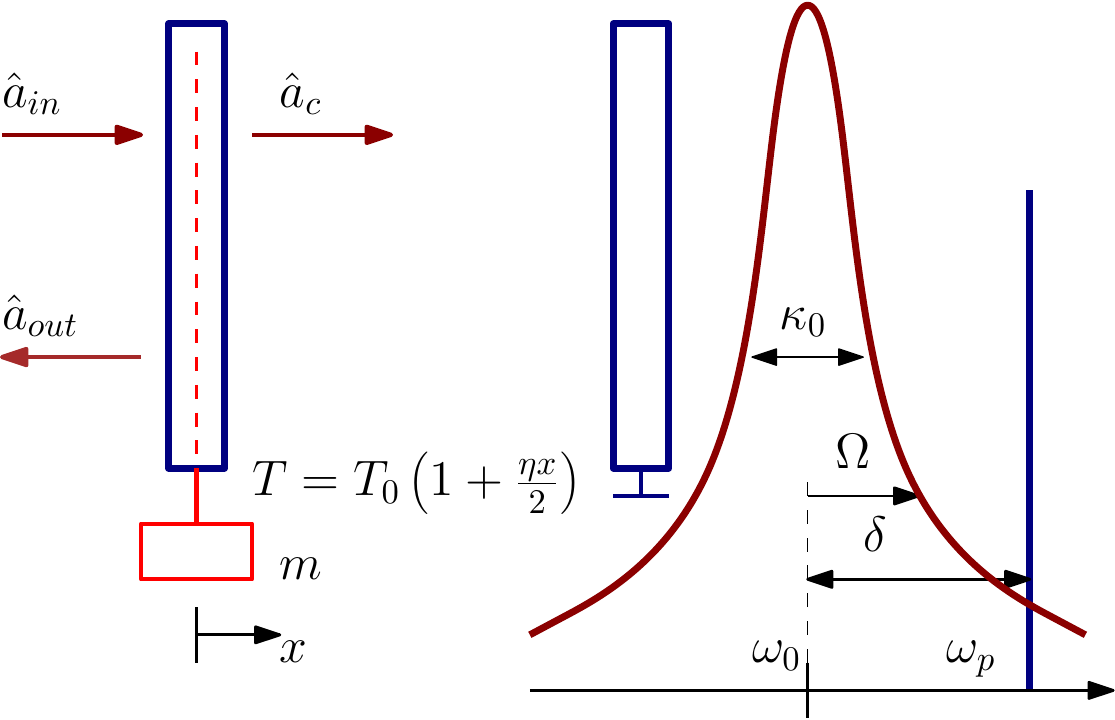}
 \caption{ Scheme  of the  Fabry-Perot cavity dissipatively coupled with displacement $x$ of free mass $m$. The coupling changes the linear amplitude transmittance of the cavity front mirror $T(x) =T_0\big[1+  \frac{\eta x}{2}\big]$ \eqref{kappa}, where  $T_0$ is the unperturbed amplitude transmission coefficient related to full width at the half maximum of the mode as $\kappa_0=T_0^2/\tau$, where $\tau$ is the round trip time of the light in the cavity. The optical mode is  pumped with the detuned coherent light ($\omega_p =\omega_0 + \delta$). }\label{DetunedDC}
\end{figure}

We consider a 1D optomechanical cavity presented on Fig.~\ref{DetunedDC}, it's optical mode with eigenfrequency $\omega_0$ is pumped with the detuned light (the pump frequency  $\omega_p =\omega_0 + \delta$) --- it is generalization of the model in \cite{16a1PRAVyMa}. The optical mode is dissipatively coupled with a mechanical system represented by a free test mass $m$. Relaxation rate $\kappa$ of the optical mode depends on the displacement $x$ of the test mass. The force of interest $F_s$ acts on the test mass and changes it's position.

We use the Hamiltonian approach to describe the dissipative coupling following \cite{ClerkRMP2010, 16a1PRAVyMa}:
 \begin{align}
 \label{Hamiltonian}
   H& =\hslash \omega_0 \hat a_c^\dag \hat a_c +\frac{\hat p^2}{2m}+H_T+H_\kappa-F_s \hat x,
 \end{align}
where $\hat a_c$ and $\hat a_c^\dag$ are the annihilation and creation operators describing the intracavity optical field, 
$\hat p$ is the momentum of the free test mass, $H_T$ describes electromagnetic continuum \cite{Marquardt09, 
AspelmeyerRMP2014},  $H_\kappa$ stands for attenuation of the pump photons and associated quantum noise. 
From the Hamiltonian \eqref{Hamiltonian} we obtain the set of corresponding equations describing time evolution of the 
optomechanical system 
\begin{subequations}
 \label{setEq}
\begin{align}
\label{ac}
 \dot{\hat a} + & \left(\frac{\kappa}{2}-i\delta \right)  \hat a = \sqrt{\kappa}\,\hat a_{in}, \\
 \label{dotx}
 \ddot{\hat x}&=i\hslash \frac{\sqrt{\kappa_0}\eta}{2m}
    \left [ \hat a^\dag \hat a_{in} - \hat a_{in}^\dag \hat a \right ]+ \frac{F_s}{m},\\
   \label{kappa}
&  \kappa=\kappa_0(1+\eta \hat x), \; \sqrt {\kappa}\simeq \sqrt{\kappa_0}\left(1+\frac{\eta}{2} \hat x\right),
 \end{align}
 Here $\hat a$ is the slow amplitude ($\hat a_c = \hat a e^{-i\omega_p t}$, see \eqref{slowA}) of the  intracavity wave, 
$\kappa$ is full width at the half maximum of the mode (relaxation rate) depending on the position $x$ of the test mass and $\eta$ 
is a constant of dissipative coupling, $\hat a_{in}$ is the slow amplitude of the input wave, see details in Appendix 
\ref{ThBa}.

These equations have to be supplied with an expression for the output amplitude $\hat a_{out}$, which can be written 
in the case of small transparency $T_0 \ll 1$ as
\begin{align}
\label{aout2}
\hat a_{out} & = -\hat a_{in}+\sqrt{\kappa}\, \hat a .
\end{align}
\end{subequations}

Below we present amplitudes as a sum of large mean and small addition values
\begin{align*}
  \hat a& \Rightarrow  A + \hat a ,\quad
    \hat a_{in}  \Rightarrow  A_{in} + \hat a_{in}  ,\quad
     \hat a_{out} \Rightarrow A_{out} + \hat a_{out} 
\end{align*}
where $A,\, A_{in}$ and $A_{out}$ are expectation values of amplitudes of the intracavity, pump and reflected waves, $\hat a_{in}$ is vacuum fluctuation wave falling on cavity, which commutator and correlator are the following
\begin{align}
  \label{comm}
  \left[\hat a_{in}(t), \hat a_{in}^\dag(t')\right] &= \delta(t-t'),\\
  \label{corr}
  \left\langle\hat a_{in}(t) \hat a_{in}^\dag(t')\right\rangle &= \delta(t-t')
\end{align}
We assume that the expectation values exceed the fluctuation parts of the operators:
\begin{align}
 \label{sa}
  A \gg  \hat a , \quad  A_{in} \gg \hat a_{in} ,\quad A_{out}\gg \hat a_{out}
\end{align}
and apply the method of successive approximations below.

Recall that from this point by $\hat a,\, \hat a_{in},\, \hat a_{out}$ we denote {\em small} slow fluctuation and signal 
additions.

We select $A_{in}=A_{in}^*$ and find steady state amplitudes
\begin{equation}
A=\frac{\sqrt{\kappa_0} A_{in}}{\frac{\kappa_0}{2}-i\delta} , \quad 
  A_{out}=A_{in}\cdot \frac{\frac{\kappa_0}{2}+i\delta}{\frac{\kappa_0}{2}-i\delta}
\end{equation}
In first order of approximation we obtain for small amplitudes  and the deviation of the test mass:
\begin{subequations}
\label{setEq3}
\begin{align}
\label{dotd}
\dot {\hat a} &+ \left(\frac{\kappa_0}{2}- i\delta\right)\hat a
    =-\frac{\eta \kappa_0}{2} A \hat x 
      +\frac{\eta \sqrt\kappa_0}{2} A_{in} \hat x+\\
      &\qquad +\sqrt{\kappa_0}\, \hat a_{in}, \nonumber\\
 \label{aout}
 &	\hat a_{out} =-\hat a_{in}+\sqrt{\kappa_0} \hat a+\frac{\eta \sqrt{\kappa_0}}{2}A\hat x, \\
 \label{dotx2}
\ddot{\hat x} &= \frac{\hat F_{lp}+F_s }{m},\\
\hat  F_{lp} & = i\hslash  \frac{\eta \sqrt{\kappa_0}}{2}
  \left [ (A^*\hat a_{in}-A\hat a^\dag_{in})- A_{in}(\hat a-\hat a^\dag)  \right ]
  \nonumber
\end{align}
\end{subequations}
Here $\hat F_{lp}$ is a light pressure force.

Below we use Fourier transform defined as
\begin{align}
 \hat a(t) &= \int\limits_{-\infty}^\infty a(\Omega) \, e^{-i\Omega t}\, \frac{d\Omega}{2\pi}
\end{align}
and by a similar way for others values, denoting Fourier transform by the same letter but without the hat. For Fourier transform of the input fluctuation operators one can derive from \eqref{comm} and\eqref{corr}:
 \begin{align}
  \label{comm1}
  \left[ a_{in}(\Omega),  a_{in}^\dag(\Omega')\right] &= 2\pi\,\delta(\Omega -\Omega'),\\
  \label{corr1}
  \left\langle a_{in}(\Omega)  a_{in}^\dag(\Omega')\right\rangle &=2\pi\, \delta(\Omega -\Omega')
\end{align}

We rewrite  Eqs. (\ref{dotd}, \ref{dotx2})  in frequency domain:
\begin{align} 
   x   =& - \frac{F_{lp}+F_s}{m  \Omega^2}, \\
   \label{Fxi}
   F_{lp}&  =\frac{i\hslash \eta  \sqrt{\kappa_0}}{2}
  \left[A^* a_{in} -A a_{in-}^\dag  - A_{in}(a - a^\dag_-)  \right],\\
  \label{alpha}
  a  =& \frac{\sqrt{\kappa_0} a_{in}}{\frac{\kappa_0}{2}- i(\Omega+\delta)}
    +\frac{\eta \sqrt \kappa_0}{2}\cdot \frac{A_{in} -\sqrt \kappa_0\,A}{\frac{\kappa_0}{2}- i(\Omega+\delta)}\cdot x.
\end{align}
Here we denote 
\begin{subequations}
    \label{notedag}
  \begin{align}
   a  &=a (\Omega),\quad a _-^\dag=a ^\dag(-\Omega) \,,\\
   a  _{in} &= a _{in}(\Omega),\quad   a _{in-}^\dag = a _{in}^\dag(-\Omega)\,.
  \end{align}
\end{subequations}

Below we present the light pressure force as a sum
\begin{align}
  F_{lp} &= F_{fl} + F_x
\end{align}
 of a fluctuation force $F_{fl}$ and a regular rigidity force $F_x$ proportional to displacement $x$, which we 
calculate in next section.

\section{Optical rigidity}

We substitute Eq.~\eqref{alpha} into the right part of \eqref{Fxi} and extract only the term $\sim x$. For the optical rigidity $K 
= -F_x /x$ one can obtain:
\begin{align}
  \label{RigidityH}
		   K & =- m\Omega_0^2
		  \cdot \frac{\delta
			 \left(\frac{\kappa_0}{2}\left[\frac{3\kappa_0}{2} -2i\Omega\right]-\delta^2\right)}{ 
		  \left[\frac{\kappa_0}{2}\right]\left(
		    \left[\frac{\kappa_0}{2}-i\Omega\right]^2 + \delta^2\right)} 
		     ,\\
	\label{Omega0}
	 \Omega_0^2 &= \frac{\hslash\, \eta^2A_{in}^2 \left[\frac{\kappa_0}{2}\right]^2}{
		  m \left(\left[\frac{\kappa_0}{2}\right]^2 + \delta^2\right)}
			= \frac{\eta^2\kappa_0 E_0}{4m\omega_p} = \frac{\eta^2W_{in}}{4m\omega_p}  
\end{align}
Here $\Omega_0^2$ is a recalculated pump (dimension of squared frequency), $E_0=\hslash \omega_p |A|^2$ is the mean energy stored in the cavity and $W_{in} = \kappa_0 E_0$ is the power of the incident wave.

Recall that in case of  dispersive coupling the optical rigidity is always unstable. For example, in case of the detuning on the right 
slope ($\delta>0$) of the resonance curve  the optical rigidity is positive but the introduced   mechanical viscosity is negative, it 
means instability (in case of the detuning on the left slope the viscosity is positive but the rigidity is negative) 
\cite{64a1eBrMi}.

In contrast, the optical rigidity \eqref{RigidityH} is more complicated compared with the rigidity based on dispersive 
coupling and one can tune both signs of the rigidity and of the viscosity by variation of relation between detuning $\delta$ and 
relaxation rate $\kappa_0$.   We can expand \eqref{RigidityH} into the Taylor series over $(-i\Omega)$  
keeping only two first terms to demonstrate it:
\begin{subequations}
 \label{Kseries}
\begin{align}
  \label{Series1}
  K_m & = - m\Omega_0^2 
		  \frac{\delta  
			\left(3\left[\frac{\kappa_0}{2}\right]^2 -\delta^2\right)}{ 
		  \left[\frac{\kappa_0}{2}\right]\left(\left[\frac{\kappa_0}{2}\right]^2 + \delta^2\right) } -\\
	\label{Series2}
	&\quad -m\Omega_0^2\frac{4 \delta  
  \left( \left[\frac{\kappa_0}{2}\right]^2 -\delta^2\right)}{
	 \left(\left[\frac{\kappa_0}{2}\right]^2 + \delta^2\right)^2}\big(- i\Omega\big)
\end{align}
\end{subequations}
It is easy to conclude that the rigidity \eqref{Series1} is positive if $\delta <0$ and $ 
3\left[\frac{\kappa_0}{2}\right]^2 >\delta^2$, whereas the viscosity \eqref{Series2} is positive if 
additionally $ \left[\frac{\kappa_0}{2}\right]^2 < \delta^2$. So we can formulate the conditions of the stable rigidity:
\begin{subequations}
\label{cond}
\begin{align}
 \label{cond1}
 \delta & < 0,\quad
 \frac{\kappa_0}{2}  < |\delta| < \sqrt 3\ \frac{ \kappa_0}{2}.
\end{align}

\begin{figure}
 \includegraphics[width=0.45\textwidth]{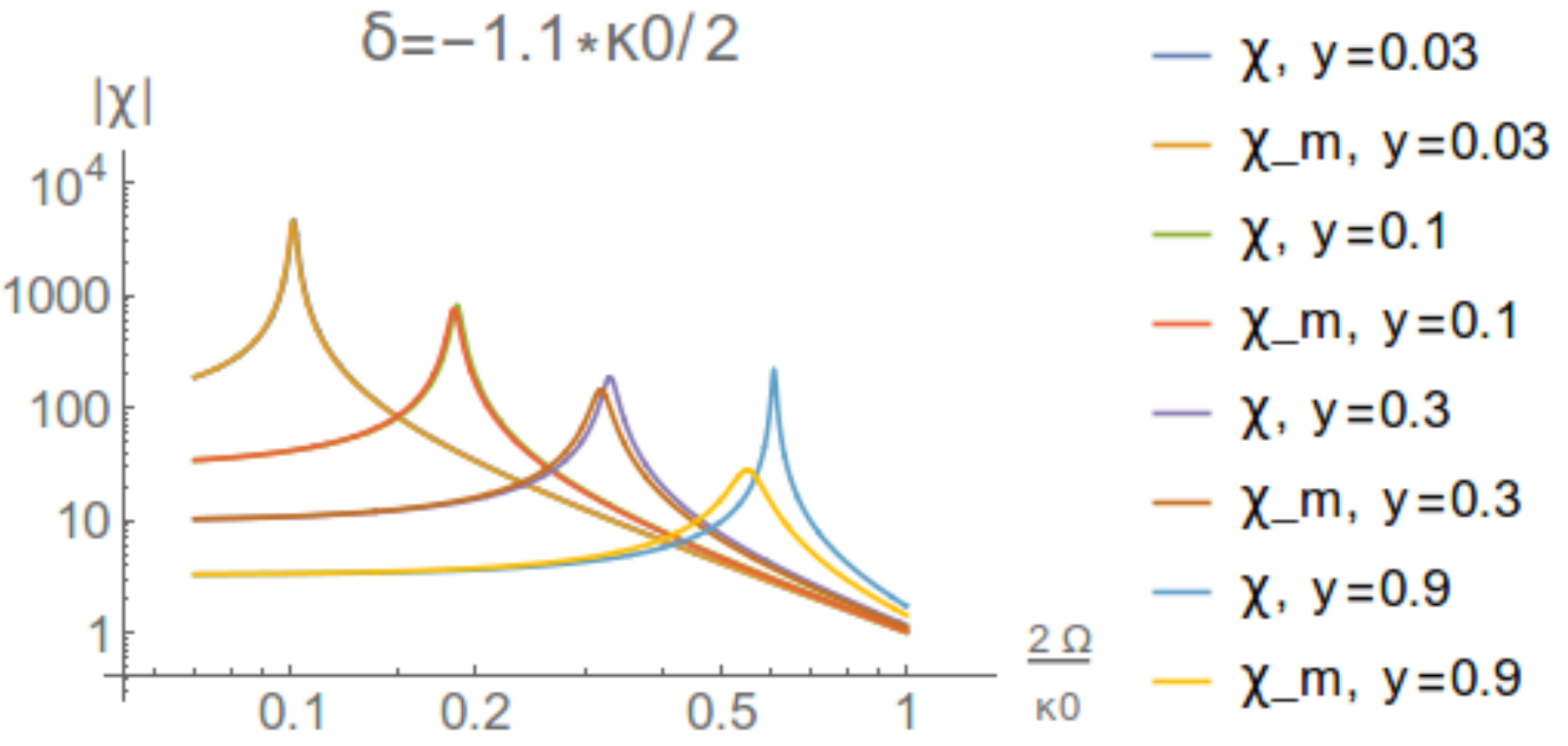}
 \includegraphics[width=0.45\textwidth]{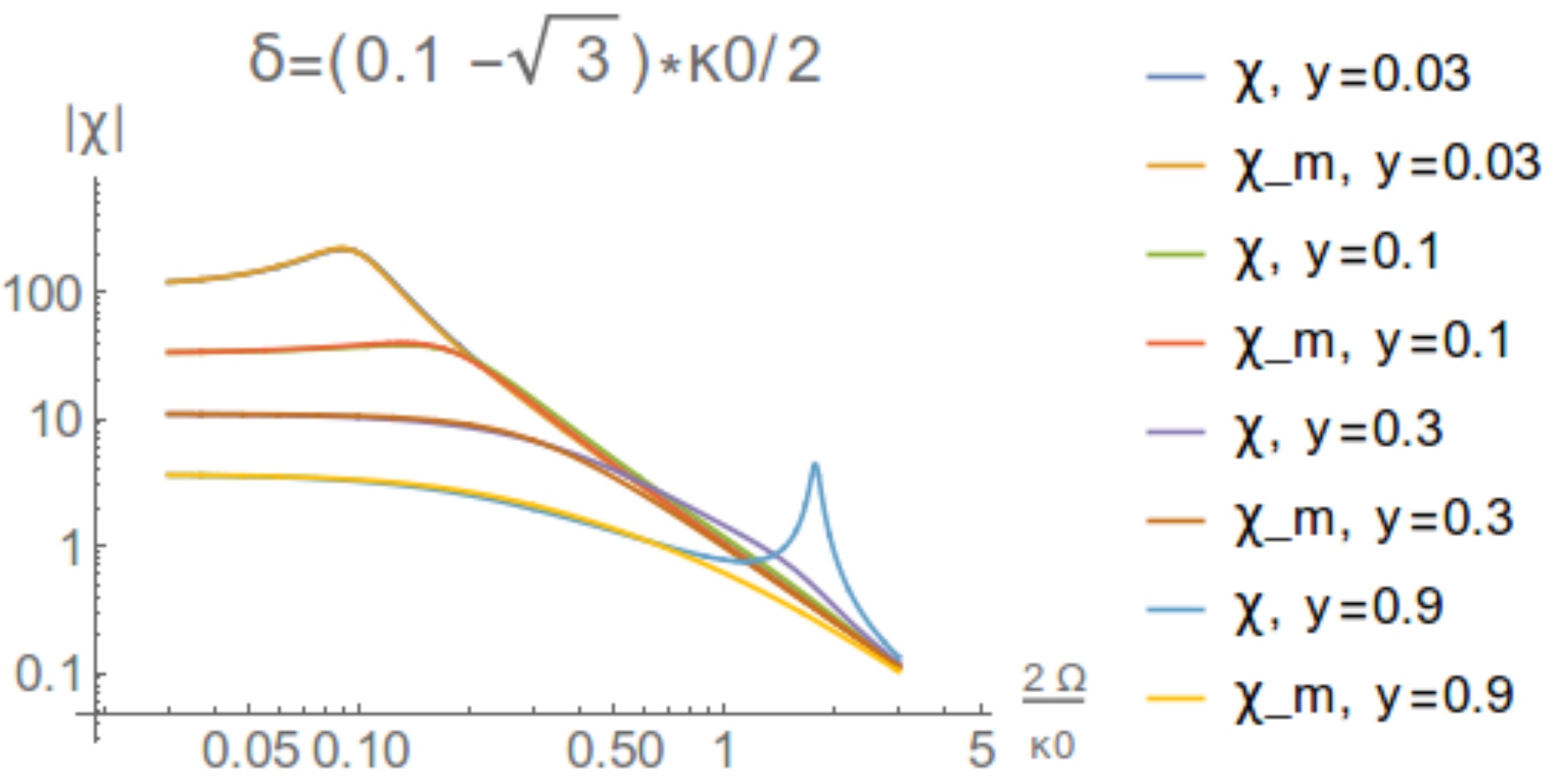}
 \caption{The plots of the susceptibilities $|\chi|$ and their approximations $|\chi_m|$ as function of frequency at detuning $\delta=-1.1\, 
\kappa_0/2$ (upper plot) and $(0.1 -\sqrt 3)\,\kappa_0/2$ as function of frequency at the different power 
parameter $y$ \eqref{y}.}\label{Sucsep}
\end{figure}

However, conditions \eqref{cond1} are a result of the approximation. For accurate consideration we apply the Routh-Hurwitz 
criterion \cite{Routh1877, Hurwitz1895, Rabinovich1984, Gopal2002} to investigate stability of the system, described by 
the susceptibility $\chi=m/(K-m\Omega^2)$, and found that the {\em accurate} conditions of stability include \eqref{cond1} plus 
one more condition on the pump
\begin{align}
\label{cond2}
 0 < \Omega_0^2< \Omega_{0max}^2,\quad &
  \Omega_{0max}^2=\frac{\kappa_0}{|\delta|}\left(\delta^2 -\left[\frac{\kappa_0}{2}\right]^2\right)\\
\label{y}  \Omega_0^2=y \Omega_{0max}^2, \quad& 0<y<1
 \end{align}
\end{subequations}
Here $y$ is dimensionless power parameter. 

Summing up, the rigidity based on dissipative coupling may be positive on {\em both left and right slopes} of the resonance curve, however, it is
stable only on left one, $\delta < 0$.

It is important that we can control characteristics of the {\em stable} rigidity. We can obtain approximation  for eigenfrequency $\Omega_m$, relaxation rate $\delta_m$ and quality factor $Q_m$ of a mechanical oscillator created by the optical rigidity using the series \eqref{Kseries} to demonstrate it:
\begin{align}
   \label{Omegam}
 \Omega_m^2 &=  \frac{\Omega_0^2|\delta|\left(3\left[\frac{\kappa_0}{2}\right]^2 - \delta^2 \right) }{ \left[\frac{\kappa_0}{2}\right]\left(\left[\frac{\kappa_0}{2}\right]^2+\delta^2\right)},\\
 \label{deltam}
    \delta_m &=  \frac{2\Omega_0^2|\delta|\left(\delta^2 -\left[\frac{\kappa_0}{2}\right]^2\right)}{(\left[\frac{\kappa_0}{2}\right]^2+\delta^2)^2},\\ 
  \label{Qm}
   Q_m &\equiv \frac{\Omega_m}{2\delta_m}= \frac{\sqrt{3\left[\frac{\kappa_0}{2}\right]^2 - \delta^2 }\,(\left[\frac{\kappa_0}{2}\right]^2+\delta^2)^{3/2}}{4\Omega_0\sqrt{|\delta|\left[\frac{\kappa_0}{2}\right]}(\delta^2-\left[\frac{\kappa_0}{2}\right]^2)}
\end{align}
Here we assumed that $\delta<0$ for stability.  

Although this consideration based on the series expansion is convenient, it is valid only for small frequency and small pump. It 
means that pump parameter $\Omega_0$ must be small. For large pump $\Omega_0$ we have to use the exact 
susceptibility  $\chi  $ instead of it's approximation $\chi_m =m/(K_m-m\Omega^2) $. On Fig.~\ref{Sucsep} we present the plots 
of $|\chi|$ and $|\chi_m|$, for the detunings $\delta$, corresponding to the stable rigidity \eqref{cond1} and different $y$. 
We see that the approximation \eqref{Kseries} gives correct results for small power parameter $y \le 0.3$ whereas for $y > 0.3$ approximation is not valid. 

Plots on Fig.~\ref{Sucsep} also show that choosing detuning $\delta$ and power parameter $y$ 
one can obtain an overdamped mechanical oscillator or an oscillator
with high quality factor. So the optical rigidity based on dissipative coupling provides a very 
promising possibility to create a mechanical system with the characteristics chosen on demand.

Introduction of the stable optical rigidity converts the free mass into the artificially created mechanical oscillator and it is interesting to estimate it's noise. Fluctuation force \eqref{Fxi} impacts on it,  it's  power spectral density $S_{Ffl}$ is equal to
\begin{align}
 S_{Ffl} &= 2\hslash m\Omega_0^2
  \frac{\left(\left[\frac{\kappa_0}{2}\right]^2 + \delta^2\right)}{\left[\frac{\kappa_0}{2}\right]^2 }\times\\
  &\qquad \times\left\{	 \big|g_- + j_-\big|^2 + \big|g_+ - j_+\big|^2\right\}.\nonumber
\end{align}
we used the definitions \eqref{Omega0}  and the formula \eqref{FflB} with the notations in Appendix~\ref{calc}. This formula can be rewritten through $\Omega_m$ using \eqref{Omegam}.

Due to action of $F_{fl}$ in equilibrium the oscillator possesses mean fluctuation energy $\mathcal E_m= m \Omega_m^2 \langle x^2\rangle$ which is convenient to characterize by mean quantum number $n_{eff}$:
\begin{align}
 \mathcal E= \hslash \Omega_m n_{eff}
\end{align}

The spectral density $S_{Ffl}$  practically does not depend on spectral frequency $\Omega$ and for high quality factor $Q_m$ \eqref{Qm}
it can be considered as a constant (white noise, not depending on $Q_m$). Consequently, mean energy $\mathcal E_m$ should increase with increase of $Q_m$. For the particular case our estimate gives 
\begin{subequations}
\begin{align}
 n_{eff} &\simeq 240,\quad \text{at}\ \delta = -0.55\,\kappa_0,\ y=0.01.
\end{align}
For these parameters 
\begin{align}
\Omega_m \simeq 0.029\, \kappa_0,\ \kappa_0 \simeq 32.37\, \Omega_0, \ Q_m \simeq 14.5. \label{Omegam2}
\end{align}
\end{subequations}

\begin{figure}
 \includegraphics[width=0.45\textwidth]{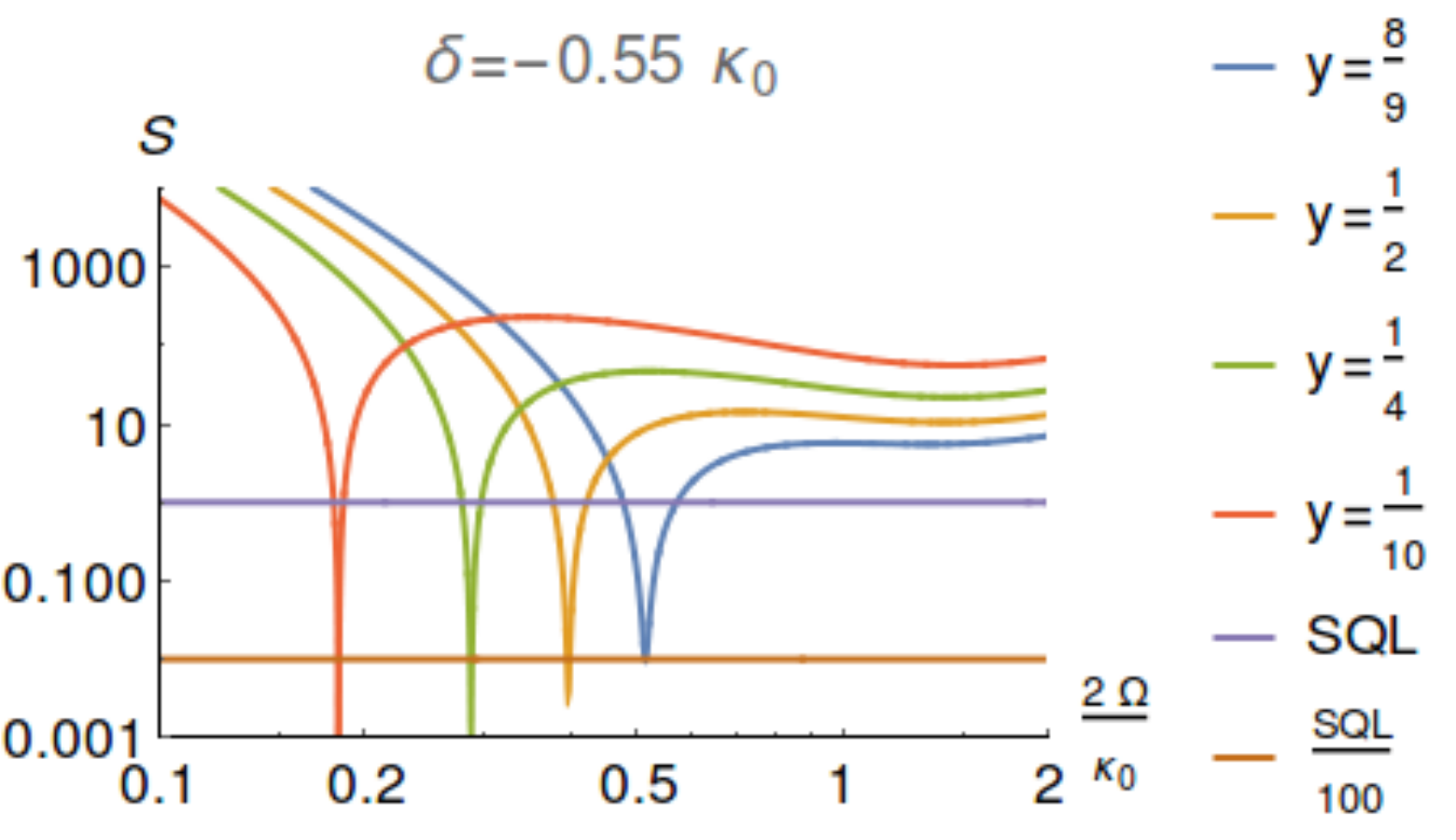}
  \includegraphics[width=0.45\textwidth]{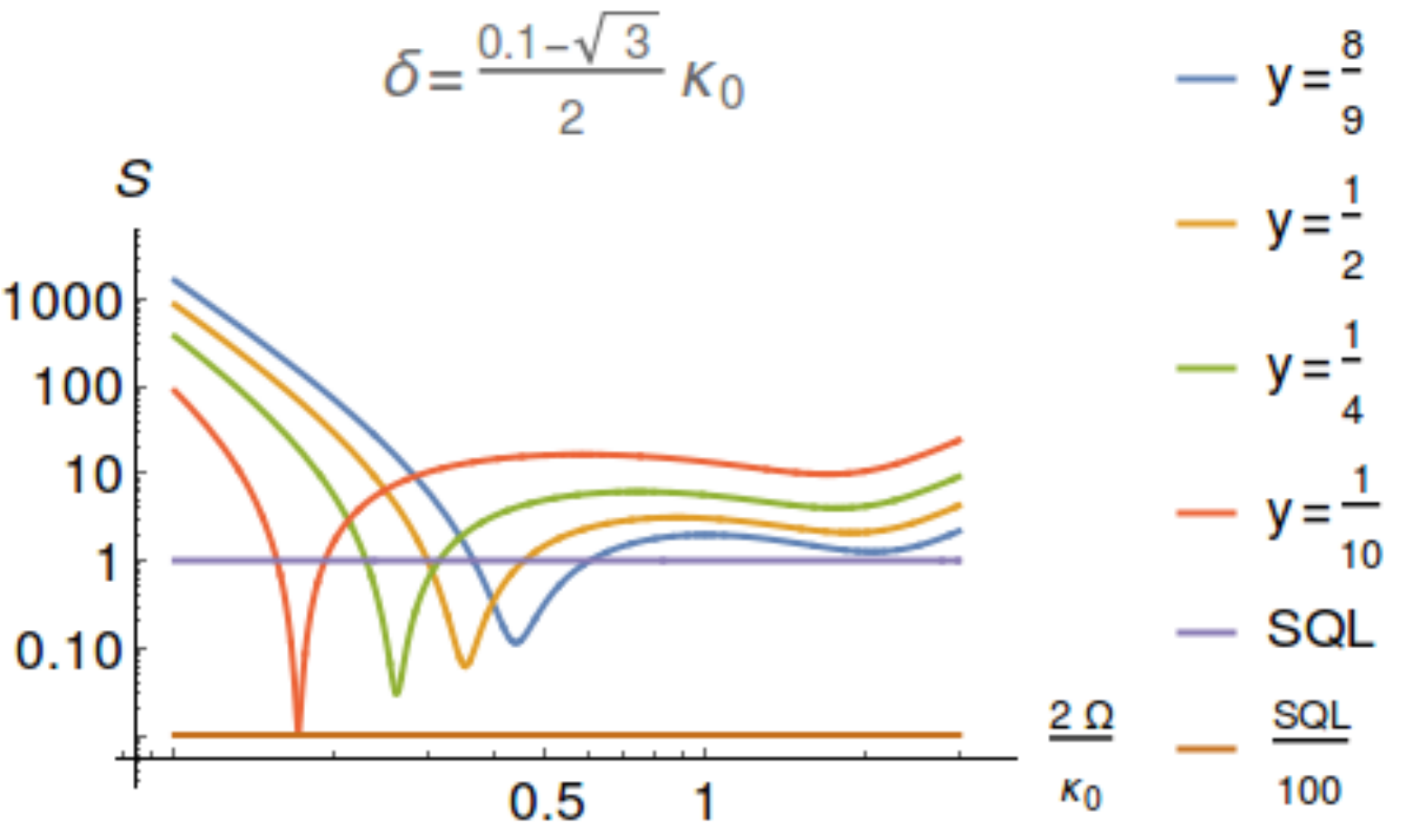}
 \caption{PSD $S_{\mathfrak f}$ as function of  frequency $\Omega$ for amplitude detection 
($\sin\theta =0$) with different power parameters $y$ \eqref{y}. On the top the detuning is $\delta =-5.5\,\kappa_0$, on the bottom --- $\delta =\frac{0.1 -\sqrt 3}{2}\,\kappa_0$.}\label{AM}
\end{figure}

\section{Detection of signal force}

Using (\ref{aout}, \eqref{alpha}) we obtain for the output amplitude in frequency domain:
\begin{align}
  \label{alphaout}
  a_{out}& = \frac{\frac{\kappa_0}{2} +i\left[\delta +\Omega\right]}{
				\frac{\kappa_0}{2} -i\left[\delta +\Omega\right]}\, a_{in}+\\
	& 	+ \frac{\eta \kappa_0 A_0 }{ 2}\cdot 
		\frac{\frac{\kappa_0}{2} +i\delta}{\frac{\kappa_0}{2} - i\delta}
		\left(\frac{1}{\frac{\kappa_0}{2} + i\delta} -
			 \frac{1}{\frac{\kappa_0}{2} - i(\delta+\Omega)}\right) \hat x ,\nonumber
\end{align}

We have to substitute  the mechanical displacement $\xi$ in frequency domain into \eqref{alphaout} with account of the rigidity \eqref{RigidityH}:
 \begin{align}
 \label{xi}
  x &= \frac{F_{fl}+F_s}{-m\Omega^2Q},\quad Q= 1 -\frac{K}{m\Omega^2}
 \end{align}
and the fluctuation force $F_{fl}$ (see details in Appendix~\ref{calc}).

We assume that the output wave is registered by a homodyne detector. Hence, we have to calculate the quadratures of the output wave. 
We define the amplitude quadrature $e_a$ andthe  phase quadrature $e_p$ inthe  output wave as following:
\begin{subequations}
 \label{eapD}
\begin{align}
 e_a &=\frac{1}{\sqrt 2}\left(\frac{\frac{\kappa_0}{2} -i\delta}{\frac{\kappa_0}{2} + i\delta}\, a_{out}        
                    +\frac{\frac{\kappa_0}{2} +i\delta}{\frac{\kappa_0}{2} - i\delta}\, a_{out-}^\dag \right) ,\\
 e_p &=\frac{1}{i\sqrt 2}\left(\frac{\frac{\kappa_0}{2} -i\delta}{\frac{\kappa_0}{2} + i\delta}\, a_{out}       
			 -\frac{\frac{\kappa_0}{2} +i\delta}{\frac{\kappa_0}{2} - i\delta}\, a_{out-}^\dag\right).
\end{align}
\end{subequations}
The calculation of the output quadratures as the functions of the  input amplitude ($a_a$) and phase ($a_p$) quadratures
\begin{align}
\label{aap}
 a_a &=\frac{a_{in} +a_{in-}^\dag}{\sqrt 2},\quad a_p =\frac{a_{in} - a_{in-}^\dag}{i\sqrt 2}
\end{align}
are presented in Appendix~\ref{calc}, the results are:
\begin{subequations}
\label{eap}
\begin{align}
 e_a &= E_{aa}a_a + E_{ap} a_p + \Phi_a f_s,\quad f_s=\frac{F_s}{\sqrt{2\hslash m \Omega^2}}\,,\\
 e_p &= E_{pa}a_a + E_{pp} a_p + \Phi_p f_s,
\end{align}
\end{subequations}
Here $f_s$ is a Fourier transform of the signal force normalized to the Standard Quantum Limit (SQL) for the free mass.
The expressions for the coefficients in \eqref{eap} are rather cumbersome and we present them using a consequence of notations
in Appendix~\ref{calc}.

In the homodyne detector we measure a quadrature $e_\theta=e_a\cos\theta +e_p\sin\theta$ in the output wave, where $\theta$ is a
homodyne angle. Sensitivity is convenient to characterize by the quadrature $e_\theta$ recalculated to the SQL:
\begin{align}
 \label{ftheta}
 \mathfrak f_\theta  = \frac{e_a\cos\theta +e_p\sin\theta}{\Phi_a\cos\theta +\Phi_p\sin\theta}
\end{align}
with the power spectral density (PSD)
\begin{align}
\label{Sf}
 S_{\mathfrak f}(\Omega) &= S_a +S_p, \\
  S_a & =\frac{\big|E_{aa}+E_{pa}\tan\theta\big|^2 }{\big|\Phi_a +\Phi_p\tan\theta\big|^2},\quad
  S_p =\frac{\big|E_{ap} +E_{pp}\tan\theta\big|^2}{\big|\Phi_a +\Phi_p\tan\theta\big|^2}\nonumber
\end{align}

Below we analyze sensitivity only for the stable rigidity (i.e. the conditions \eqref{cond} are valid).

\begin{figure}
 \includegraphics[width=0.45\textwidth]{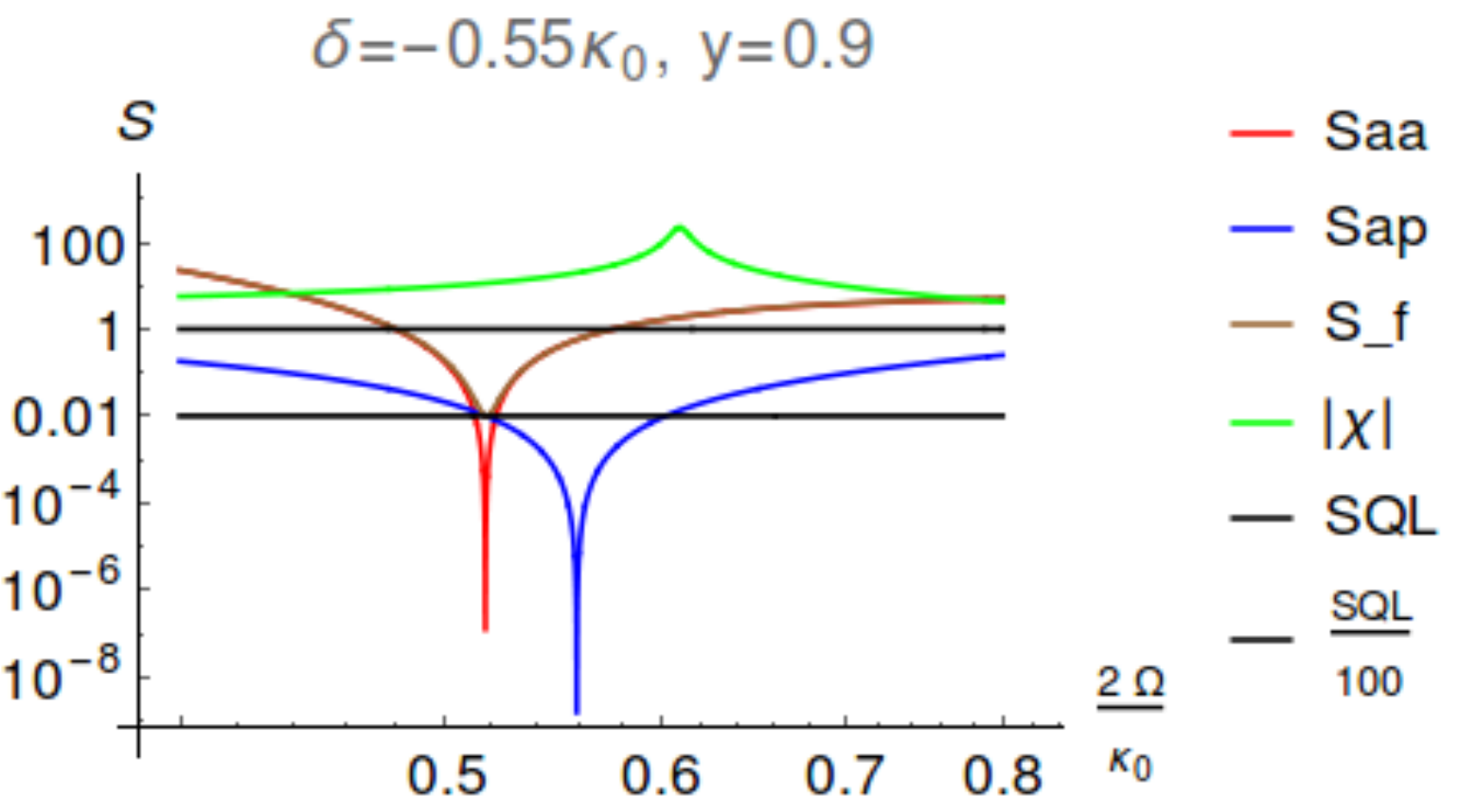}
 \caption{Amplitude detection. Presentation of the contributions by the different terms $S^A_{a}$ and $S^A_{p}$ \eqref{SfAM} into the power spectral density $S_{\mathfrak f}$  for the particular power parameter $y=0.9$ and the detuning $\delta = - 0.55\, \kappa_0$. The  plot of the susceptibility $|\chi|$ is also presented.}\label{AMb}
\end{figure}

\subsection{Amplitude detection}

Amplitude detection is simpler to realize in experiment as compared with homodyne one. Formally it corresponds to 
$\sin\theta=0$ in the formulas \eqref{Sf}:
\begin{align}
\label{SfAM}
 S_{\mathfrak f}(\Omega) &= S^A_{a} + S^A_{p},\quad  S^A_{a} =\frac{\big|E_{aa}\big|^2}{\big|\Phi_a\big|^2},\ 
    S^A_{p} =\frac{\big|E_{ap}\big|^2}{\big|\Phi_a\big|^2}
\end{align}
We obtain that even amplitude detection allows to surpass the SQL (i.e. $S_\mathfrak f<1$) by more 
than 100 times. Choosing the pump parameter $y$ one can vary both spectral frequency and range of the SQL 
surpassing  --- plots on Fig.~\ref{AM} demonstrates it. 

The top plots on Fig.~\ref{AM} demonstrate the  SQL overcoming by about 1000 times but in the narrow bandwidth. In contrast,  the bottom 
plots demonstrate the more modest SQL overcoming by about 100 times but in the wider bandwidth. Note that the pump power on the top plots is about 10 times lesser than on the bottom ones (with the same pump parameter $y$).

Analysis shows that the SQL surpassing takes place when the coefficient $E_{aa}$ has minimum due to the compensation of the shot noise term $\sim \beta_+$ and the backaction term $\sim \Omega_0^2$ --- see \eqref{E_aa} in Appendix~\ref{calc}. The same compensation takes place for the coefficient $E_{ap}$ in \eqref{E_ap}, but on slightly different spectral frequency $\Omega$. 

For demonstration we present on Fig.~\ref{AMb} the contributions of the different terms $S^A_{a}$ and $S^A_{p}$ of the 
spectral density $S_{\mathfrak f}$ \eqref{SfAM} for the particular pump parameter $y=0.9$ and the detuning $\delta=- 
0.55\, \kappa_0$. The plot of the susceptibility is also presented (it has different dimension) in order to show that 
the mentioned compensation takes place on frequencies {\em different} from  frequency $\Omega_m$ of mechanical 
resonance. 

\begin{figure}
 \includegraphics[width=0.45\textwidth]{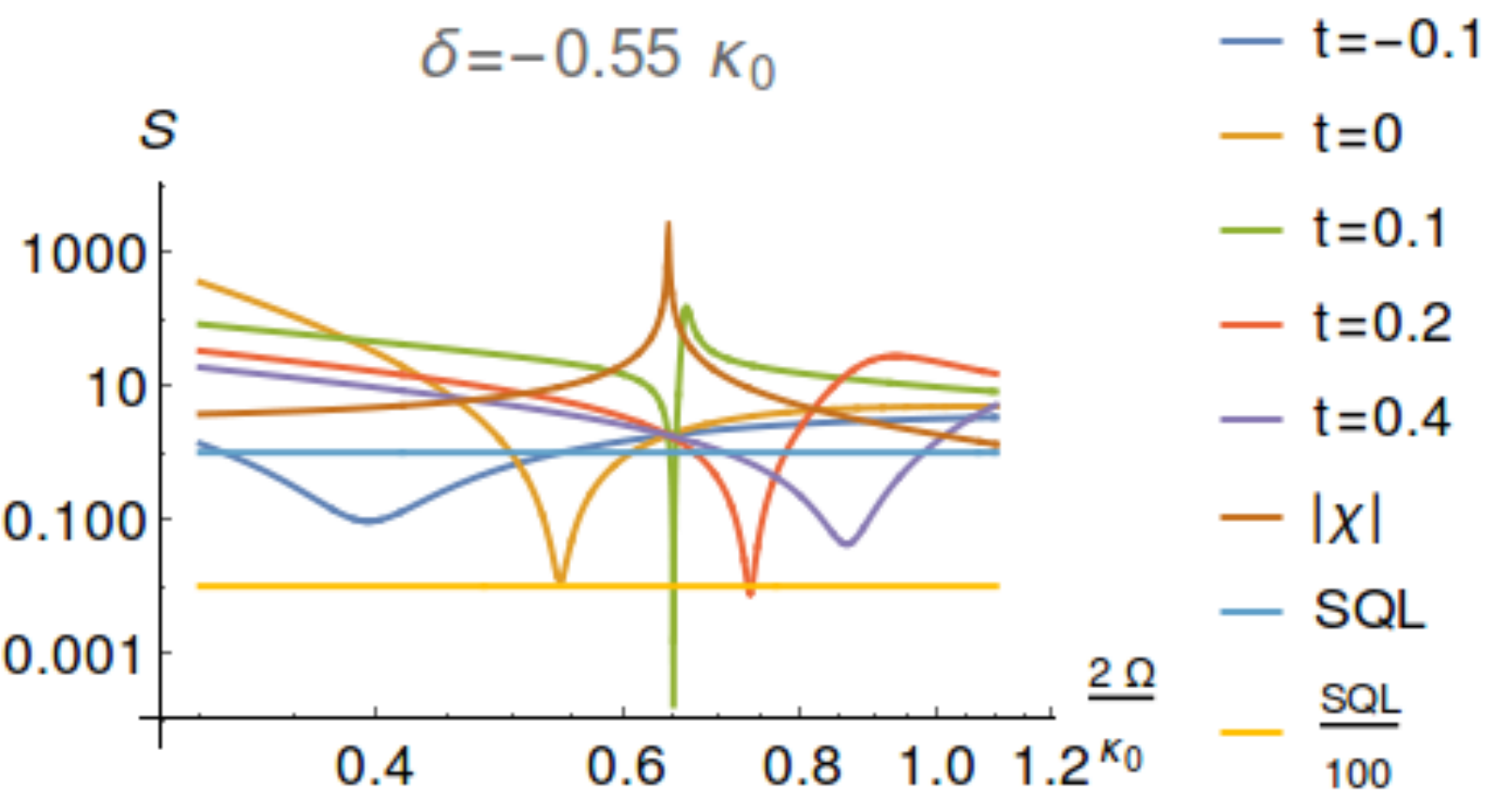}
 \includegraphics[width=0.45\textwidth]{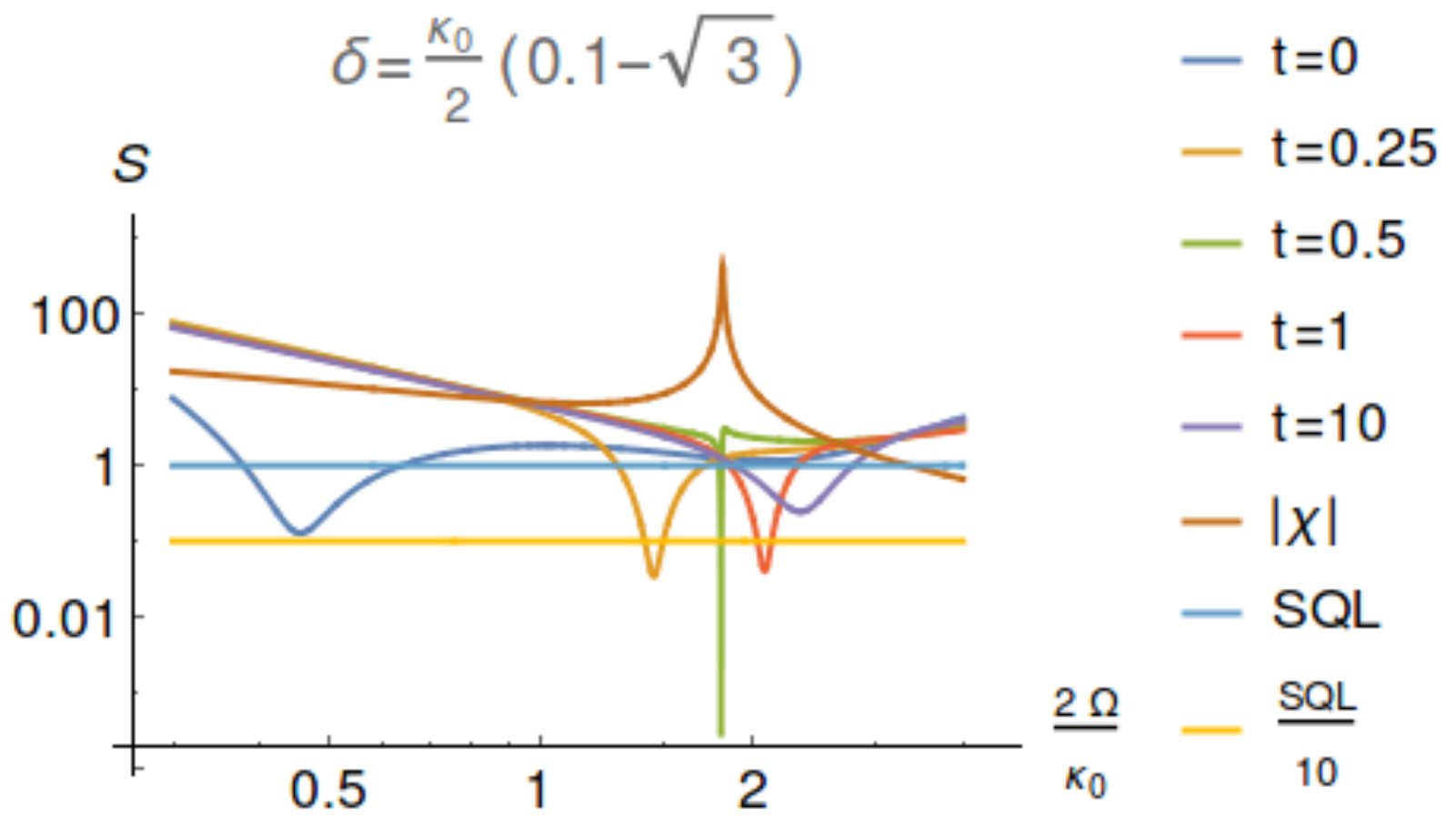}
 \caption{The PSD $S_{\mathfrak f}$ as a function of  frequency $\Omega$ for homodyne detection 
 with the  constant power parameter $y= 8/9$ \eqref{y} at the different homodyne angles $t= \tan\theta$. On the top the detuning is $\delta =-5.5\,\kappa_0$, on the bottom --- $\delta =\frac{0.1 -\sqrt 3}{2}\,\kappa_0$. The  plot of the susceptibility $|\chi|$ is also presented.}\label{HD}
\end{figure}

\subsection{Homodyne detection}

In this case we have the homodyne angle $\theta$ as an additional degree of freedom which provides a possibility  to control the sensitivity. Indeed, even at the constant pump we can change the PSD by the tuning of the homodyne angle. As shown on Fig.~\ref{HD} the PSD has a minimum at frequency $\Omega_{min}$ and it's width $\Delta \Omega$ (where the SQL is surpassed, i.e. $S_\mathfrak f<1$) can be shifted and changed. 

The plots on Fig.~\ref{HD} demonstrate that frequency $\Omega_{min}$ grows with increase of the homodyne angle $\theta$, whereas the bandwidth $\Delta \Omega$ initially decreases until $\Omega_{min}<\Omega_m$ and increases when $\Omega_{min}>\Omega_m$. Note that if $\Omega_{min} \simeq \Omega_m$ we have the most strong minimum of the PSD but in the very narrow bandwidth.

Detailed analysis shows that for the particular plots on top of Fig.~\ref{HD} the amplitude part $S_a$ makes the main contribution into the PSD \eqref{Sf}. The minimum of the PSD (practically the minimum of $S_a$) takes place when the shot noise term $\sim (\beta_+ - \beta_-^* \tan\theta )$ and the backaction noise term $\sim(E_{aa1} + E_{pa1}\tan\theta)$ compensate each others --- for details see formulas \eqref{Eap0} in Appendix~\ref{calc}. The plot of the susceptibility is also presented (it has different dimension) in order to show that 
the mentioned compensation takes place on frequencies {\em close} to  frequency $\Omega_m$ of the mechanical 
resonance.

\section{Model of Dissipative Coupling based on Michelson-Sagnac interferometer}

\begin{figure}
 \includegraphics[width=0.45\textwidth]{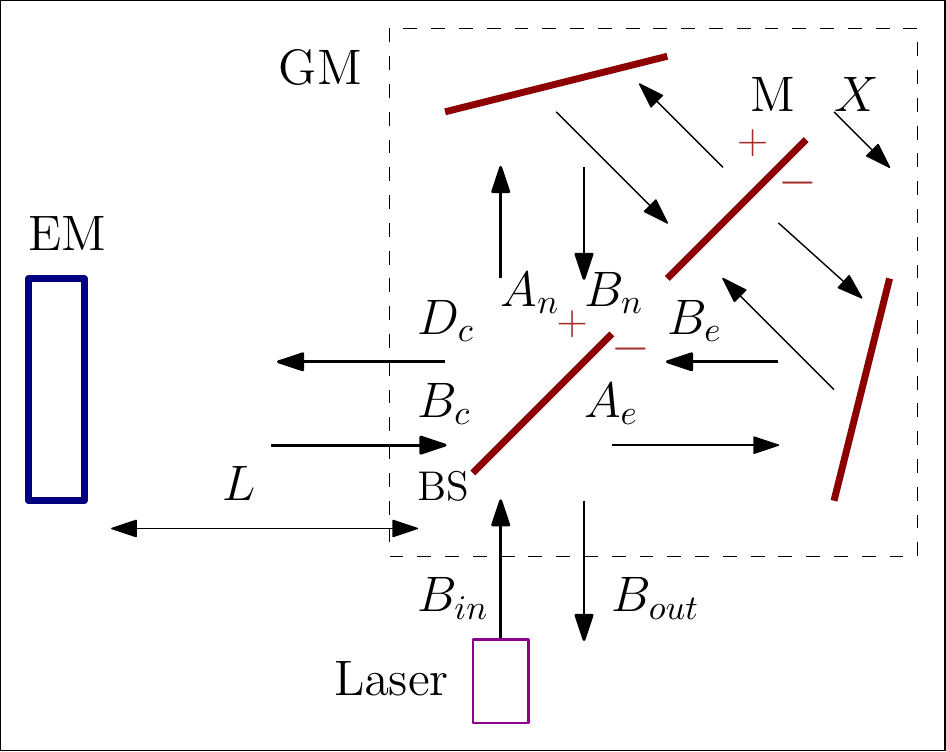}
 \caption{MSI with the input non-balanced beam splitter ($T_{bs}\ne R_{bs}$) and the movable partially reflective 
mirror $M$ with mass $m$. It plays the role of the input GM  of the cavity pumped by the laser detuned from 
resonance.}\label{MSI}
\end{figure}

For realization of dissipative coupling without dispersive one we consider a MSI first 
suggested in \cite{XuerebPRL2011}. In the Fabry-Perot cavity, shown in Fig.~\ref{MSI}, the MSI 
plays the role of the input generalized  mirror (GM). Here we present the generalized model with the non-balanced beam splitter with amplitude transmittance $T_{bs}$ and  reflectivity $R_{bs}$ and  the partially reflecting mirror $M$ with transmittance $T$ and reflectivity $R$. We assume that the GM size is smaller than the distance $L$ between  the not movable 
beam splitter and the end mirror so both amplitude transmittance $\mathbb T$ and reflectivity $\mathbb R$ of the GM depend on 
position $X$ of movable mirror $M$ with mass $m$ and do not depend on spectral frequency. 

We start from the boundary conditions on the beam splitter:
\begin{subequations}
 \label{TRbs}
 \begin{align}
   A_e &= T_{bs}B_c - R_{bs}B_{in},\ A_n = R_{bs}B_c + T_{bs} B_{in},\\ 
   B_{out} &= T_{bs} B_{n} - R_{bs} B_e ,\ D_c = T_{bs}B_e + R_{bs} B_{n},
   \end{align}
\end{subequations}
where $B_{in},\ B_{out},\ B_c,\  D_c,\ A_e,\ A_n,\ B_e,\ B_n$ are the complex amplitudes of the incident and reflected waves on the beam splitter --- see the notations on Fig.~\ref{MSI}. The boundary conditions on the mirror M give
\begin{subequations}
 \label{TRM}
   \begin{align}
   B_e & =  - R A_e  e^{2ik\ell_e}+ T A_ne^{ik(\ell_e+\ell_n)},\\
   B_n & =  RA_n e^{2ik\ell_n} +T A_e e^{ik(\ell_e+\ell_n)}
\end{align}
\end{subequations}
where $k=\omega_p/c$ is a wave vector and $c$ is the speed of light, $k\ell_e$ ($k\ell_n$) is the accumulated phase of
the light traveling between the beam splitter and the mirror $M$ through the   east (north) arm. 

We define reflectivity and transmittance of the GM as
\begin{align}
 D_c &=  \mathbb T B_{in} + \mathbb R_\rhd B_c, \quad B_{out} = \mathbb T B_c + \mathbb R_\lhd B_{in}.                                                                                        
\end{align}
Using \eqref{TRbs} and \eqref{TRM} one can derive
\begin{subequations}
  \label{TTRR}
  \begin{align}    
   \label{TT}
    \mathbb T &= e^{i\phi_+} \left\{2R R_{bs}T_{bs}\cos \phi_- 
	  - T\Delta_{bs}   \right\},\\
    \mathbb R_\rhd  &= Re^{i\phi_+}\times\\
	&\qquad \times 
	\left\{ \Delta_{bs} \cos  \phi_- - i  \sin  \phi_- 
	    + \frac{2T T_{bs}R_{bs}}{R}\right\} ,\nonumber\\
    \mathbb R_\lhd &=  - Re^{i\phi_+}\times\\
     &\qquad \times \left\{ \Delta_{bs}\cos \phi_- +i\sin \phi_- 
	    + \frac{2 TT_{bs}R_{bs}}{R} \right\},\nonumber\\
	&\phi_\pm = k\big(\ell_e \pm \ell_n\big),\quad \Delta_{bs} =R_{bs}^2 - T_{bs}^2
 \end{align}
 \end{subequations}
It is obvious that the sum phase $\phi_+$ does not depend on the  displacement $X$ of the mirror M, but the phase difference $\phi_-$ does 
depend. Below we present the displacement $X=x_0 +x$  as a sum of the constant mean value $x_0$ and the small addition $x$ so that 
$\phi_-=\phi_0 + 2 k x$ and expand  reflectivity  and transmittance of the GM in a series over $x$.

One can easy derive that for the realization of pure dissipative coupling (but not combination of dissipative and dispersive ones) we must have the relative derivatives of $\mathbb T$, $\mathbb R_\rhd$ and $\mathbb R_\lhd$ over $\phi_-$ to be real. Calculations give:
\begin{subequations}
  \begin{align}
  \label{diffTT}
   \frac{\partial_{\phi_-} \mathbb T}{\mathbb T}& = 
      \frac{-2R R_{bs}T_{bs}\sin \phi_-}{2R R_{bs}T_{bs}\cos \phi_-  - T\Delta_{bs} },\\
	\label{diffRR}
    \frac{\partial_{\phi_-} \mathbb R_\rhd }{\mathbb R_\rhd }& =
      \frac{-R \Delta_{bs}\sin  \phi_- - i R \cos  \phi_- }{ 
	  R\Delta_{bs}\cos  \phi_- - i R \sin  \phi_-  + 2T T_{bs}R_{bs}} .
    \end{align}
 We see that the relative derivative \eqref{diffTT} is real at any combination of the parameters. In order to have the real derivative \eqref{diffRR} we have two possibilities:

 a) the balanced beam splitter ($\Delta_{bs}=0$) and the perfectly reflective mirror M ($T=0$);  this case was analyzed in \cite{XuerebPRL2011, 16a1PRAVyMa};
 
 b) the non-balanced beam splitter and the  partially transparent mirror M ($T\ne 0$) --- in this case we have to choose 
$\phi_-=\phi_0$, where $\phi_0$ is the solution of equation  
   \begin{align}
   \label{condphi}
	 \cos  \phi_0 &= \frac{- R\Delta_{bs}}{2T T_{bs}R_{bs}},\\
   	 \label{TT3}
    \mathbb T_0 = \mathbb T|_{\phi_-=\phi_0} & = -e^{i\phi_+} \frac{\Delta_{bs}}{T},\\
    |\mathbb R_0| = |\mathbb R_\rhd|_{\phi_-=\phi_0}& = \frac{\sqrt{(2T_{bs}R_{bs})^2 -R^2}}{T},\\
    \left. \frac{\partial_{\phi_-} \mathbb T}{\mathbb T}\right|_{\phi_-=\phi_0}& 
      =\frac{2R T R_{bs}T_{bs}\sin \phi_0}{\Delta_{bs}} =\frac{R|\mathbb R_0|}{|\mathbb T_0|}.
 \end{align}
 \end{subequations}
 The cavity should have high finesse. Hence, for $|\mathbb T_0|\ll 1$ one have to have $|\Delta_{bs}| \ll T$. 
So assuming $ e^{i\phi_+}= - 1$ we obtain in first order  approximation over $x$
\begin{align}
\label{TR0}
  \mathbb T&= \mathbb T_0\left( 1  + \frac{R |\mathbb R_0|}{|\mathbb T_0|}\,2k x +\dots \right) , \\
  \mathbb R_\rhd &= \mathbb R_0 \left(1 - \frac{R |\mathbb T_0|}{|\mathbb R_0|}\, 2kx +\dots\right).
\end{align}

We see that for the realization of dissipative coupling with the partially transparent mirror M we should choose the correct angle 
$\phi_0$) (i.e. the constant displacement $x_0$). It is important that we can choose the parameters of the GM on demand by variation of the beam 
splitter parameters $(R_{bs}, \ T_{bs})$. 

The small displacement $x$ of the mirror $M$ from the mean position $x_0$ provides modulation of the relaxation rate of 
the Fabry-Perot interferometer:
\begin{subequations}
 \label{MSIeta}
\begin{align}
 \kappa &=\kappa_0 (1+\eta x), \quad  \kappa_0  = \frac{|\mathbb T|^2}{\tau}, \quad
      \tau=\frac{2L}{c}\\
    \eta &= 4k \frac{R |\mathbb R_0|}{|\mathbb T_0|} \,.
\end{align}
\end{subequations}
It is easy to demonstrate that  all equations for this optomechanical system are the same as derived in Sec.~\ref{model}.

It is important that on the example of the considered interferometer as the GM we can demonstrate the peculiar property of a light pressure force in an optomechanical system with dissipative coupling. Indeed,  using the notations on Fig.~\ref{MSI} we can write a ponderomotive force acting on the mirror M:
\begin{align}
 \label{F1b}
 F = & 2\hslash k R^2\big(| A_n|^2-|A_e|^2 \big)=\\
    & = 4\hslash k R^2\left( B_c  B_{in}^* +  B_c^*  B_{in}\right).
\end{align}
Here in the last equation we used the input-output relation \eqref{TRbs} putting $R_{bs}=T_{bs}$.
Recall that for dispersive coupling the ponderomotive force is just proportional to the square of the amplitude of the intracavity wave. In
contrast, for the optomechanical system with dissipative coupling the force is proportional to the {\em cross product} of 
 the incident $ B_{in}$  and the inside $ B_{c}$ amplitudes as it follows from \eqref{F1b}.  The light 
pressure force depends on phase difference between $ B_{in}$  and  $\tilde B_{c}$, so it can be also called as the {\em 
interferometric} pressure. It is precisely this property provides the additional possibilities for the realization of 
the stable rigidity. We would like to pay attention on resemblance between formulas \eqref{F1b} and \eqref{Fxi}, obtained in frame of the Hamiltonian approach.

Note that the realization of the stable optical rigidity  was proposed \cite{TarabrinPRA2013} and elegantly demonstrated 
\cite{SawadskyPRL2015,Corbitt2018} for a similar scheme with {\em alone} MSI presented on Fig.~\ref{MSI}, without any
cavity and without focusing attention on dissipative or dispersive coupling is used.
In contrast, in the scheme analyzed in this paper the stable optical rigidity is a property of cavity with dissipative 
coupling and  we formulated the conditions when MSI is a generalized mirror with dissipative coupling (but not a combination 
of dissipative and dispersive ones).

\section*{Conclusion}

 We analyzed  the optical rigidity based on dissipative coupling and formulated the conditions \eqref{cond} of the {\em stable} optical rigidity. Recall that using dispersive coupling one can get only the {\em unstable} rigidity \cite{99a1BrKhPLA, 01a1KhPLA}.

The rigidity based on dissipative coupling may be positive on {\em the  both left and right slopes} of the resonance curve (but 
stable only on the left one, $\delta < 0$), 
whereas the positive (unstable) rigidity in case of dispersive coupling takes place on {\em the right} slope only.

We show that physical reason of stability of the rigidity based on dissipative coupling is interference between the input and intracavity waves, originating the more complicated dependence of the light pressure force as compared with dispersive coupling.

We have shown that dissipative coupling can be realized in  the MSI with the partially transparent 
mirror M --- it is the generalization of the previous results \cite{XuerebPRL2011,16a1PRAVyMa} for the perfectly 
reflecting mirror M. It provides the possibility to use a thin membrane \cite{Arcizet2006b, Zwickl2008, Borkje10} as 
the mirror M with extra small mass $m$ for the experimental realization. 

For the estimation we assume:
\begin{subequations}
 \label{param}
 \begin{align}
 m &= 10^{-8}\, \text{g},\quad k= \frac{2\pi}{\lambda} , \quad \lambda=10^{-6}\,\text{m}, \\
 & W_{in}=10^{-4}\, \text{W},\quad R^2=0.7,\quad |\mathbb T|^2=10^{-4}
\end{align}
\end{subequations}

Using \eqref{Omega0}, \eqref{Omegam} we obtain the estimations of the power parameter $\Omega_0$ and  the mechanical eigenfrequency $\Omega_m$:
\begin{subequations}
\begin{align}
 \label{est}
 \Omega_0 &= \sqrt\frac{4kW_{in}}{mc|\mathbb T|^2}\simeq 92\cdot 10^3\, \text{rad/s},\\
 \Omega_m &\simeq 86\cdot 10^3\, \text{rad/s}.
\end{align}
\end{subequations}
In the last estimation we put $\delta=-0.55\, \kappa_0$.

It means the possibility to create a mechanical nano-oscillator with the eigenfrequency in the  range of hundreds kHz  from a free mass and the stable optical rigidity. The fluctuation light pressure force is a source of the  excitation of the oscillator, we show that in equilibrium the  mean quantum number $n_{eff}$ of such oscillator may be about $200$. This estimate corresponds to coherent pump, however, for specially tuned squeezed pump mean quantum number $n_{eff}$ can be smaller.

\section*{Acknowledgments}

Authors acknowledges support from  Russian Science Foundation (Grant No. 17-12-01095).

\appendix

\section{Dissipation Description}\label{ThBa}
In this Appendix we present the detailed description of the Hamiltonian \eqref{Hamiltonian} and the derivation of the equations \eqref{setEq} for the field $\hat a_c$ inside cavity and the mechanical coordinate $\hat x$.

We write the Hamiltonians $H_T,\ H_{\kappa}$ in form
\begin{align}
\label{HinitTB}
   H_\kappa &= -i\hslash\sqrt\frac{\kappa\, \Delta \omega}{2\pi}\,
      \sum_{q=1}^\infty\left(\hat a_c^\dag\hat b_q -\hat b_q^\dag \hat a_c\right)\, ,\\
   H_T & =       \sum_{q=1}^\infty\hslash \omega_q\, \hat b_q^\dag\hat b_{q}\,,
\end{align}
Here we present a thermal bath as infinite number of the oscillators with the annihilation and creation operators $\hat b_q, \ \hat b_q^\dag $, $q$ is integer number, frequencies $\omega_q$ of these oscillators are separated by $\Delta \omega= \omega_q -\omega_{q-1}$, we hold in mind that below we  put
\begin{align}
   \label{limit}
 \Delta \omega\to 0
\end{align}
 The commutators and correlators are
\begin{align}
   \label{commb}
  \big[\hat b_q, \hat b_{q'}^\dag\big]= \delta_{q q'},\quad
   \big\langle\hat b_q \hat b_{q'}^\dag\big\rangle= \, \delta_{q q'}\,.
\end{align}
(The temperature of the bath is assumed to be zero.)
We write down the movement equations:
\begin{subequations}
  \label{TB1}
  \begin{align}
   \label{TBhata}
    \dot{\hat a}_c &=\frac{1}{i\hslash}\big[\hat a_c,H\big] = - i \omega_0 \hat a_c
      - \sqrt\frac{\kappa\, \Delta \omega}{2\pi } \sum_{q=1}^\infty \hat b_q ,\\
      \dot{\hat b}_q &= - i \omega_q \hat b_q
	 + \sqrt\frac{\kappa\, \Delta \omega}{2\pi} \hat a_c.
  \end{align}
 Introducing the slow amplitudes
  \begin{align}
   \label{slowA}
	\hat a_c & = \hat a e^{-i\omega_p t},\quad
	 \hat b_q\Rightarrow \hat b_q e^{-i\omega_q t}
 \end{align}
  we get
  \begin{align}
   \label{dotadotb}
	\dot{\hat a}(t) - i\delta\, \hat a &=   - \sqrt\frac{\kappa \, \Delta \omega}{2\pi}
	    \sum_{q=1}^\infty \hat b_q e^{i(\omega_p-\omega_q)t},\\
\label{dothatb}
	 \dot{\hat b}_q &=
	 \sqrt\frac{\kappa\, \Delta \omega}{2\pi}\, \hat a e^{-i(\omega_p-\omega_q)t}.
\end{align}
\end{subequations}
We substitute the formal solution of \eqref{dothatb} for $\dot{\hat b}_q$ into \eqref{dotadotb} using method of successive approximations based on \eqref{limit}:
\begin{subequations}
  \label{TB}
  \begin{align}
  \label{TBhatbq}
    \hat b_q &= \hat b_q(0) + \sqrt\frac{\kappa\, \Delta \omega}{2\pi}
      \int\limits_0^t \hat a_c(t') e^{-i(\omega_p-\omega_q)t'}\, dt' ,\\
    \label{TBhata2}
    \dot{\hat a} & - i\delta\, \hat a = -  \sum_{q=1}^\infty \left( \sqrt\frac{\kappa\, \Delta \omega}{2\pi}\,
      \hat b_q(0) e^{i(\omega_p-\omega_q)t}- \right.\\
      \label{TBhata3}
      &\quad \left. - \frac{\kappa\, \Delta \omega}{2\pi} \,\int\limits_0^t \hat a_c(t')\,  e^{i(\omega_p-\omega_q)(t-t')}\,dt'
      \right)=\\
      \label{dissdi}
      	& = \sqrt{\kappa} \, \hat a_{in} - \frac{\kappa}{2} \, \hat a\, .
     \end{align}
     \end{subequations}
   Below we present the details of the derivation \eqref{dissdi}.

   In the further calculations in  the limit \eqref{limit} we replace the sum by the integral using  the rule
   \begin{align}
   \label{rule}
    \Delta\omega \sum_{q=1}^\infty \Rightarrow \int\limits_0^\infty  d\omega_q \, .
   \end{align}

   Using \eqref{commb} we calculate the commutator \eqref{comm} for $a_{in}$ defined in \eqref{TBhata2}
   \begin{subequations}
    \begin{align}
    \label{ain2}
	\hat a_{in}(t) &\equiv \sqrt \frac{\Delta\omega}{2\pi} \sum_{q=1}^\infty \hat b_q(0) e^{i(\omega_p-\omega_q)t},\\
	\big[\hat a_{in}(t),\hat a_{in}^\dag(t')\big] &
	\equiv \frac{\Delta\omega}{2\pi} \sum_{q=1}^\infty e^{i(\omega_p-\omega_q)(t-t')}=\\
	& = \delta(t-t')\, ,
      \end{align}
 Here $\delta(t)$ is the Dirac delta function. By the similar calculation we obtain the correlator \eqref{corr} assuming that the thermostat oscillators are in the main state: 
 \begin{align}
  \left\langle b_q b_{q'}^\dag\right\rangle&  =\delta_{kk'},\quad 
    \left\langle b_q^\dag b_{q'}\right\rangle =0
 \end{align}
where $\delta_{kk'}$ is the Kronecker delta.
  
 We calculate the  term \eqref{TBhata3} using the rule \eqref{rule}:
    \begin{align}
	 & \int\limits_0^t \hat a_c(t')\left[ \sum_{q=1}^\infty\frac{\kappa\,\Delta\omega}{2\pi} \,
	    e^{i(\omega_p-\omega_q)(t-t')} \right] dt'=\\
	 &  = \int\limits_0^t \hat a_c(t')\,  \frac{\kappa}{2\pi}\left[\int\limits_0^\infty \, e^{i(\omega_p-\omega_q)(t-t')}d\omega_q \right] dt'=\\
	      &\qquad  = \frac{\kappa}{2\pi} \,\int\limits_0^t \hat a_c(t')\, 2\pi \,\delta(t-t') dt'= \frac{\kappa}{2} \, \hat a_c(t)\,.
  \end{align}
\end{subequations}
 From \eqref{HinitTB}  we get for the mechanical coordinate using the definition \eqref{kappa}
   \begin{align}
   \ddot{\hat x}&=\frac{i\hslash\eta}{2m}\sqrt\frac{\kappa_0\, \Delta \omega}{2\pi}\,
      \sum_{q=1}^\infty\left(\hat a_c^\dag\hat b_q -\hat b_q^\dag \hat a_c\right)+ \frac{F_s}{m}.
    \end{align}
    Using the definition \eqref{ain2} we obtain \eqref{dotx}.

\section{Calculations of the output quadratures}\label{calc}

Here we present the details of calculations for the output quadratures. Here we use the following notations in order to compact the formulas below:
\begin{subequations}
 \label{psi}
 \begin{align}
  \psi &= \frac{\kappa_0}{2}-i\delta,\quad \psi^* =  \frac{\kappa_0}{2}+i\delta,\\
  \label{Psi}
  \Psi&= \frac{\kappa_0}{2}-i\delta-i\Omega, \quad
    \Psi^* = \frac{\kappa_0}{2}+i\delta+i\Omega,\\
   \Psi_-&= \frac{\kappa_0}{2}-i\delta+i\Omega, \quad  \Psi^*_- = \frac{\kappa_0}{2}+i\delta-i\Omega\,,
 \end{align}
\end{subequations}
For the fluctuation part of the light pressure force $F_{fl}$ in frequency domain we get using \eqref{Fxi} and the notations \eqref{notedag}:
\begin{align}
 \label{Ffl}
 F_{fl} &= \frac{i\hslash A_0\eta \sqrt \kappa_0}{2}
 \left\{ 
    \frac{a_{in-}^\dag}{\psi}  
	- \frac{a_{in}}{\psi^*} 
		- \frac{a_{in-}^\dag}{\Psi^*_-}  
		  + \frac{a_{in}}{\Psi} \right\},\\
 g_+ &\equiv\frac{\kappa_0}{4}\left(\frac{1}{\psi }
			  + \frac{1}{\psi^*}\right),\quad
	 g_- \equiv\frac{\kappa_0}{4i}\left(\frac{1}{\psi }
			  - \frac{1}{\psi^*}\right) \, ,\\
	j_+ &\equiv\frac{\kappa_0}{4}\left(\frac{1}{\Psi }
			  + \frac{1}{\Psi_-^*}\right)\,,\quad
	 j_- \equiv\frac{\kappa_0}{4i}\left(\frac{1}{\Psi }
			  - \frac{1}{\Psi_-^*}\right).
\end{align}
and express it through the quadratures \eqref{aap} of the input wave
\begin{align}
\label{FflB}
 F_{fl} &= \sqrt 2 \hslash A_0 \eta \left\{
	 -a_a \big(g_- + j_-\big) + a_p\big(g_+ - j_+\big)\right\}
\end{align}
Substituting it into \eqref{xi} and then into \eqref{alphaout} using \eqref{aap} we obtain:
\begin{align}
 a_{out} &=\frac{\Psi^*}{\Psi}\, a_{in}
   +  \frac{\eta \kappa_0 A_0 }{ 2}\cdot \frac{\psi^*}{\psi}
	 \left(\frac{1}{\psi^*} - \frac{1}{\Psi}\right) \times\\
	 &\quad \times \frac{\sqrt 2 \hslash A_0 \eta}{-m\Omega^2 Q} \left\{
	 -a_a \big(g_- + j_-\big) + a_p\big(g_+ - j_+\big)\right\}. \nonumber
\end{align}
Then we substitute it into the definitions \eqref{eapD} and after simple but awkward calculations we finally obtain the coefficients in \eqref{eap}:
\begin{subequations}
\label{Eap0}
\begin{align}
 \label{E_aa}
 E_{aa} & = \beta_+ + E_{aa1},\\
   & E_{aa1} = \frac{ 2\Omega_0^2}{\Omega^2Q} \left(g_+  - j_+\right) \big(g_- + j_-\big),\\
 \label{E_ap}
 E_{ap} & = - \beta_-^* + E_{ap1},\\
   & E_{ap1}=\frac{ 2\Omega_0^2}{\left(\Omega^2Q\right)} \left(-g_+  + j_+\right)\big(g_+ - j_+\big),\\
 \Phi_a & =  \sqrt{\frac{4\Omega_0^2}{ \Omega^2}}\left(\frac{-g_+  + j_+}{Q}\right),\\
 E_{pa} & = \beta_-^* -   E_{pa1},\\
   &E_{pa1} = \frac{ 2\Omega_0^2}{ \Omega^2Q} \left(g_-  + j_-\right) \big(g_- + j_-\big),\\
 E_{pp} & = \beta_+ + E_{pp1},\\
   & E_{pp1}=\frac{ 2\Omega_0^2}{ \Omega^2Q} \left(g_-  + j_-\right) \big(g_+ - j_+\big),\\
 \Phi_p & = \sqrt{\frac{4\Omega_0^2}{ \Omega^2}}\left(\frac{g_- + j_-}{Q}\right).
\end{align}
\end{subequations}
where 
\begin{subequations}
\label{gpm}
\begin{align}
\beta_+ &=\frac{ 1}{ 2}\left(\frac{\psi\Psi^*}{\psi^*\Psi} + \frac{\psi^*\Psi_-}{\psi\Psi_-^*}\right),\\
\beta_- &=\frac{ 1}{ 2i}\left(\frac{\psi\Psi^*}{\psi^*\Psi} - \frac{\psi^*\Psi_-}{\psi\Psi_-^*}\right)
\end{align}
\end{subequations}
and $\Omega_0^2$ is the normalized pump \eqref{Omega0}.

\end{document}